\documentclass[prd,aps,preprintnumbers, showpacs, nofootinbib,superscriptaddress,notitlepage]{revtex4-1}
\usepackage{mathrsfs}
\usepackage{amsfonts}
\usepackage{amsmath}
\usepackage{array}
\usepackage{verbatim}
\usepackage{epsfig}
\usepackage{graphicx}

\newcommand{\beq}{\begin{eqnarray}}
\newcommand{\eeq}{\end{eqnarray}}

\newcommand{\tr}{{\rm tr}  }
\newcommand{\nn}{\nonumber \\}

\begin{document}

\begin{flushright}
YITP-16-1
\end{flushright}

\title{Probing the Small-$x$ Gluon Tomography in Correlated Hard Diffractive Dijet
Production in DIS}

\author{Yoshitaka Hatta}
\affiliation{Yukawa Institute for Theoretical Physics, Kyoto University, Kyoto 606-8502, Japan}

\author{Bo-Wen Xiao}
\affiliation{Key Laboratory of Quark and Lepton Physics (MOE) and Institute
of Particle Physics, Central China Normal University, Wuhan 430079, China}

\author{Feng Yuan}
\affiliation{Nuclear Science Division, Lawrence Berkeley National
Laboratory, Berkeley, CA 94720, USA}

\begin{abstract}
We investigate the  close connection between the quantum phase space Wigner
distribution of small-$x$ gluons and the color dipole scattering amplitude, and propose to 
study it experimentally in the hard diffractive dijet production at the planned electron-ion 
collider. 
The angular correlation between the nucleon recoiled momentum and the dijet
transverse momentum will probe the nontrivial correlation in the phase space 
Wigner distribution.
This experimental study will not only provide us with three-dimensional tomographic pictures 
of gluons inside high energy proton, but also give a unique and interesting signal for the 
small-$x$ dynamics with QCD evolution effects.
\end{abstract}

\maketitle

{\it Introduction.}
There have been strong interests in hadron physics community \cite{Boer:2011fh, AbelleiraFernandez:2012cc, Accardi:2012qut} to
explore the partonic structure of the nucleon, in particular, aiming at
a tomography picture from which we can image the partons in 
three-dimensional fashion. This can provide fruitful and detailed information
on the sub-atomic structure of the baryonic building blocks of the universe,
and deepen our understanding of the strong interaction facts
in constructing the fundamental particles. Among these tomography
distributions, the so-called quantum phase space Wigner distributions \cite{Ji:2003ak, Belitsky:2003nz}
of partons have been reckoned as the mother distributions
of all, since they ingeniously encode all quantum information of how partons are distributed inside hadrons. 

The key question now is to find experimental probes to measure
these distributions. The goal of this paper is to pioneer this direction, by
pointing out that we can have access to the gluon Wigner distributions
at small-$x$. The proposed new observables will stimulate further
developments from both experiment and theory sides for the planned
electron-ion colliders (EIC).
In general, it is believed that the parton Wigner distributions normally 
are not directly measurable in high energy scatterings. Due to the 
uncertainty principle, Wigner distributions, which are not positive 
definite, are only quasi-probabilistic. As we will demonstrate later 
in this Letter, one can use the diffractive dijet production (or more 
complicated processes), which has been a subject of study in the 
small-$x$ physics and the generalized parton distribution approach~\cite{Nikolaev:1994cd,
 Bartels:1996ne, Bartels:1996tc, Diehl:1996st, Braun:2005rg,Rezaeian:2012ji},
to directly probe the Fourier transform of the gluon Wigner distribution 
at the EIC. 

The phase space distributions~\cite{Mueller:1999wm} of quarks 
and gluons are often used in small-$x$ literatures, and they are 
believed to be possibly related to Wigner distributions~\cite{Ji:2003ak}, 
although the exact connection was not known. We will show that the gluon Wigner
distributions at small-$x$ can be simplified and written as the Fourier transform of well-known
impact parameter dependent dipole amplitudes, which helps us to build intimate connections
to small-$x$ factorization framework developed in the last few decades. 
This will not only provide the motivation to pursue the gluon Wigner
distributions in the future EIC, but also prompt further studies to
investigate non-trivial correlations in the small-$x$ dipole scattering
amplitude. The latter has become one of the most important elements
of the phenomenological studies in heavy ion collisions and deep 
inelastic scatterings~\cite{Gelis:2010nm, Kovchegov:2012mbw}.

One of the nontrivial phenomena is the angular correlation between
the traverse momentum of the produced dijet and the recoiled momentum
of the nucleon, which provides vital information
on the gluon Wigner distributions. It is important to emphasize that this correlation can help
us test and measure the unique feature of angular correlations between impact parameter and
dipole size predicted by small-$x$ evolutions.

The rest of the paper is organized as follows. We first introduce
the gluon Wigner distributions and take the small-$x$ limit, which
can be connected to the dipole scattering amplitudes.
We then explore the small-$x$
dynamics by invoking the analytical solution to the BFKL 
equation~\cite{BFKLeq} to show there exist
nontrivial correlation in these gluon Wigner distributions. Last,
we apply these results to demonstrate that we will be able to
observe these novel correlations in the future EIC. Finally, we
summarize our paper in the end.

{\it Gluon Wigner Distributions at Small-$x$.}
The parton Wigner distributions are introduced to describe the
quantum phase space distributions of partons inside the nucleon.
They unify the two common languages of
transverse momentum dependent and the generalized parton
distributions in parton distributions framework.

We focus on the gluon Wigner distributions. Similar study can
be done for the quark part. The gluon Wigner distributions
are defined through the following matrix elements,
\begin{eqnarray}
xW_g^T(x,\vec{q}_\perp;\vec{b}_\perp)=\int\frac{d\xi^-d^2\xi_\perp}{(2\pi)^3P^+}\int \frac{d^2\Delta_\perp}{(2\pi)^2} e^{-ixP^+\xi^--iq_\perp\cdot \xi_\perp}
\left\langle P+\frac{\Delta_\perp}{2}\left|F^{+i}\left(\vec{b}_\perp+\frac{\xi}{2}\right)F^{+i}
\left(\vec{b}_\perp-\frac{\xi}{2}\right)\right|P-\frac{\Delta_\perp}{2}\right\rangle\,,
\end{eqnarray}
where $F^{\mu\nu}$ represents the field strength tensor, $x$ and $q_\perp$ for the longitudinal
momentum fraction and the transverse momentum for the gluon, $\vec{b}_\perp$ for the
coordinate space variable. The above Wigner distributions not only provide us with the 
momentum distribution of the corresponding parton, but also give the average location 
of that parton w.r.t. the center of target hadron. In the following, we will work in the 
small-$x$ limit where we can simplify the above expression and relate it to the dipole 
scattering amplitudes commonly used in the small-$x$ factorization approach. 
The Fourier transform of the Wigner distribution w.r.t. the impact parameter $b_\perp$
is also referred as the generalized transverse momentum dependent (GTMD) gluon 
distribution \cite{Meissner:2009ww,Lorce:2013pza}.

In Ref.~\cite{Dominguez:2010xd, Dominguez:2011wm}, it has been demonstrated that 
TMD gluon distributions are related to small-$x$ unintegrated gluon distributions. 
The Weizs\"{a}cker-Williams (WW) and the dipole gluon distribution used in 
small-$x$ formalism correspond to two gauge invariant but topologically 
different operator definitions. In order to pursue deeper connections between 
Wigner distributions and small-$x$ impact parameter dependent gluon 
distributions, we first use the dipole gluon distribution as an example, 
and we will comment on the case of the WW gluon distribution in the 
end. Following the convention in Ref.~\cite{Bomhof:2006dp}, 
we write down the GTMD dipole gluon distribution as
\begin{equation}
xG_{ \textrm{DP}  }(x,q_{\perp }, \Delta_\perp)=2\int \frac{\text{d}\xi^{-}\text{d}%
^{2}\xi _{\perp }e^{-iq_{\perp }\cdot \xi _{\perp }-ixP^{+}\xi ^{-}}}{\left(
2\pi \right) ^{3}P^{+}}\left\langle  P+\frac{ \Delta_\perp}{2}\left|\text{Tr}\left[ F^{+i}\left(\xi/2\right)
\mathcal{U}^{\left[ -\right] \dagger }F^{+i}\left( -\xi/2\right) \mathcal{U}^{\left[ +%
\right] }\right] \right| P-\frac{ \Delta_\perp}{2}\right\rangle\,, \label{gtd}
\end{equation}
 where $\mathcal{U}^{[\pm]}$ are the future/past-pointing U-shaped Wilson lines which make the operator gauge invariant. 
Its Fourier transform 
$\int \frac{d^2 \Delta_\perp}{(2\pi)^2} e^{i\Delta_\perp\cdot b_\perp} xG_{ \textrm{DP}  }(x,q_{\perp }, \Delta_\perp)$
can be identified as the Wigner distribution $xW_g^{T}(x, q_\perp, b_\perp)$. 
Following similar derivation used in Ref~\cite{Bomhof:2006dp, Belitsky:2002sm, Dominguez:2011wm} 
in the small-$x$ limit, one can show that Eq.~(\ref{gtd}) reduces to
\begin{equation}
xG_{ \textrm{DP}  }(x,q_{\perp }, \Delta_\perp)=\frac{2N_{c}}{\alpha_s}\int
\frac{d^{2}R_{\perp } d^{2}R_{\perp }^{\prime }}{(2\pi)^4}e^{iq_{\perp }\cdot \left(
R_{\perp }-R_{\perp }^{\prime }\right) +i \frac{\Delta_\perp}{2}\cdot(R_\perp+R_\perp^\prime)} \left(\nabla _{R_{\perp }}\cdot
\nabla _{R_{\perp }^{\prime }}\right)\frac{1}{N_{c}}\left\langle\text{Tr}\left[
U\left( R_{\perp }\right) U^{\dagger }\left( R_{\perp }^{\prime }\right)
\right]\right\rangle_x\,, \label{op}
\end{equation}
where we can recognize the impact parameter dependent dipole amplitude. Let us define its double Fourier transform
\begin{equation}
\frac{1}{N_{c}}\text{Tr}\left[
U\left( b_{\perp }+\frac{r_\perp}{2}\right) U^{\dagger }\left( b_{\perp }-\frac{r_\perp}{2} \right)\right] \equiv \int d^2 q_\perp d^2 \Delta_\perp e^{-iq_\perp \cdot r_\perp -i \Delta_\perp\cdot b_\perp} \mathcal{F}_x(q_\perp, \Delta_\perp)\,.\label{g2}
\end{equation}
 Then we can succinctly write $xG_{ \textrm{DP}  }(x,q_{\perp }, \Delta_\perp)=(q_\perp^2 -\Delta_\perp^2/4)\frac{2N_c}{\alpha_s}  \mathcal{F}_x(q_\perp, \Delta_\perp) $.
 Setting $r_\perp=0$ in the above expression, we obtain the normalization condition for $\mathcal{F}_x(q_\perp, \Delta_\perp)$ as
$\int d^2 q_\perp d^2 \Delta_\perp e^{ -i \Delta_\perp\cdot b_\perp}  \mathcal{F}_x(q_\perp, \Delta_\perp) =1$.

{\it Gluon Tomography Induced by Small-$x$ Dynamics.}
In the small-$x$ literature, the dipole scattering amplitudes have been
widely used to describe the relevant processes such as the inclusive DIS,
inclusive hadron production in $pA$ collisions. In these calculations,
the dipole amplitude is averaged over the azimuthal angle of impact parameter $\vec{b}_\perp$. 
This is because most of the observables discussed in phenomenology so far are not sensitive to 
the $\vec{b}_\perp$-dependence.  On the other hand, there have been
theoretical investigations \cite{Lipatov:1985uk,Navelet:1997tx,Navelet:1997xn} on the correlation between $\vec{b}_\perp$
and the dipole size $\vec{r}_\perp$, which eventually leads to  nontrivial correlations
between $\vec{b}_\perp$ and the transverse momentum $\vec{q}_\perp$ 
in the gluonic Wigner distributions. 
In the following, as a simple example, we illustrate how these correlations 
 are generated at small-$x$ in the BFKL approximation. This provides an intuitive picture of
small-$x$ gluon distributions in the nucleon.

Let us consider the dipole scattering amplitude off a dipole $x_\perp$ (quark at $\vec{x}_\perp/2$, antiquark at $-\vec{x}_\perp/2$) evolved up to rapidity $Y=\ln 1/x$. 
Define the dipole T-matrix in impact parameter space as
$\frac{1}{N_c}\left\langle \tr\,  U\left(b_\perp+\frac{r_\perp}{2}\right) U^\dagger\left(b_\perp-\frac{r_\perp}{2}\right)\right\rangle_x = 1-T(r_\perp,b_\perp, Y)$.
 In the BFKL approximation, $T$ is given by \cite{Lipatov:1985uk,Navelet:1997tx,Navelet:1997xn}
\beq
T(r_\perp,b_\perp, Y)&=&2\pi\alpha_s^2\sum_n\int\frac{d\nu}{(2\pi)^3}
\frac{(1+(-1)^n)\left(\nu^2+\frac{n^2}{4}\right)}{\left(\nu^2+\left(\frac{n-1}{2}\right)^2\right)
\left(\nu^2+\left(\frac{n+1}{2}\right)^2\right)}e^{\chi(n,\nu)Y} \nn
&& \qquad \times
\int d^2\omega_\perp E^{1-h,1-\bar{h}}(b_\perp+\frac{r_\perp}{2}-\omega_\perp,b_\perp-\frac{r_\perp}{2}-\omega_\perp)E^{h,\bar{h}}
(\frac{x_\perp}{2}-\omega_\perp,-\frac{x_\perp}{2}-\omega_\perp)\,, \label{bfkl}
\eeq
where $E^{h,\bar{h}}(a-\omega,b-\omega)=(-1)^n\left(\frac{z_{ab}}{z_{a\omega}z_{b\omega}}\right)^h \left(\frac{\bar{z}_{ab}}{\bar{z}_{a\omega}\bar{z}_{b\omega}}\right)^{\bar{h}}$ is the BFKL holomorphic eigenfunction with $h=\frac{1-n}{2}+i\nu$ and $\bar{h}=\frac{1+n}{2}+i\nu$. $\chi(n,\nu)\equiv \frac{2\alpha_sN_c}{\pi} \left[\psi (1) -\textrm{Re}\psi (\frac{|n|+1}{2}+i\nu)\right]$ is the BFKL characteristic function. In the high energy limit, the $n=0$ part of the solution gives the leading contribution.
Using the saddle point approximation around $\nu=0$ and taking the limit $x_\perp \ll b_\perp,r_\perp$, we can cast Eq.~(\ref{bfkl}) into
\beq
T(r_\perp,b_\perp, Y)\approx \frac{\alpha_s^2 |\rho|}{\sqrt{\pi} } \frac{\ln\frac{16}{|\rho|}}
{\left(\frac{7}{2}\bar{\alpha}_s\zeta(3)Y\right)^{3/2}}\exp\left(4\bar{\alpha}_sY\ln 2 -\frac{\ln^2 \frac{16}{|\rho|}}{14\bar{\alpha}_s \zeta(3)Y}\right)\ , \label{he}
\eeq
where 
\beq
|\rho|^2\equiv \frac{x_\perp^2 r_\perp^2}{\left(b_\perp+\frac{r_\perp}{2}-\frac{x_\perp}{2}\right)^2 \left(b_\perp-\frac{r_\perp}{2}+\frac{x_\perp}{2}\right)^2} \approx \frac{x_\perp^2 r_\perp^2}{b_\perp^4+\frac{r_\perp^4}{16}-\frac{b_\perp^2r_\perp^2}{2}\cos 2\phi_{b r}}\,.
\eeq
Clearly, one sees that there is nontrivial angular correlation between $b_\perp$ and $r_\perp$ contained in Eqs.~(\ref{bfkl}) and (\ref{he}).
When $b_\perp$ is parallel to $r_\perp$, the scattering is stronger than the case when 
$b_\perp$ is perpendicular to $r_\perp$. This is a known phenomenon, see for example 
Ref.~\cite{Kopeliovich:2007fv}. Such a correlation is expected to survive near the 
nonlinear saturated regime. Indeed, away from the BFKL saddle point, the saturation 
momentum $Q_s$ is defined by the condition $T(r_\perp=1/Q_s,b_\perp)=const.$ which leads to
\beq
\frac{1}{|\rho|^2}\approx \left.\frac{b_\perp^4+\frac{r_\perp^4}{16}-\frac{b_\perp^2r_\perp^2}{2}\cos 2\phi_{br}}{x_\perp^2r_\perp^2} \right|_{r_\perp=1/Q_s}\sim e^{\frac{\chi(\gamma_s)}{\gamma_s}Y} \,,
\eeq
 where $\gamma_s=\frac{1}{2}+i\nu_s=0.628$. If we look for a solution in the regime $b_\perp \gg r_\perp \simeq1/Q_s$, we find
 \beq
 Q_s^2 \sim \frac{x_\perp^2}{b_\perp^4}e^{\frac{\chi(\gamma_s)}{\gamma_s}Y} +\frac{\cos 2\phi_{br}}{2b_\perp^2}.
 \eeq
This is consistent with the numerical study of the nonlinear small-$x$ evolution 
(e.g., the Balitsky-Kovchegov evolution \cite{Balitsky:1995ub, Kovchegov:1999yj}) 
in Ref.~\cite{GolecBiernat:2003ym,Berger:2011ew} where it was observed that the 
angular correlation exists even when $b_\perp$ and $r_\perp$ are of the same order.
These features should be a guiding principle when building saturation models with angular correlations.

The elliptic ($\sim\cos 2\phi$) angular correlation can be seen also in the momentum space. 
After averaging over the angular orientation of the target dipole $x_\perp$, we find that the 
Fourier transform of $T(r_\perp,b_\perp, Y)$ w.r.t. $b_\perp$ and $r_\perp$ is
 \begin{eqnarray}
 \mathcal{T}(q_\perp,\Delta_\perp, Y)& \equiv& \int \frac{d^2r_\perp d^2 b_\perp}{(2\pi)^4}  e^{iq_\perp \cdot r_\perp+i\Delta_\perp \cdot b_\perp} T(x_\perp, r_\perp, b_\perp, Y) \nonumber \\
&\simeq& \frac{  \alpha_s^2 x_\perp}{ (2\pi)^2 \Delta_\perp^3} \frac{e^{4\bar{\alpha}_sY\ln 2}}
{\left(\frac{7}{2}\bar{\alpha}_s\zeta(3)Y \pi \right)^{3/2}} \int_0^{\pi/2} d\theta J_0\left(\frac{\sin \theta \Delta_\perp  x_\perp}{2}\right)K_{0}\left(\frac{\cos \theta \Delta_\perp x_\perp}{2} \right) \notag\\
  && \times \int_0^1\frac{d\alpha}{\alpha^2(1-\alpha)^2}\,_2F_1\left(\frac{3}{2},\frac{3}{2},1,
  -\frac{|\vec{q}_\perp+(1/2-\alpha) \vec{\Delta}_\perp|^2}{\Delta_\perp^2 \alpha(1-\alpha)} \right),
\end{eqnarray}
in the high energy limit. Depending on relative size of $q_\perp$ and $\Delta_\perp$, 
$\mathcal{T}(q_\perp,\Delta_\perp, Y)$ could have sizable angular correlations with only 
even harmonics (it is not hard to show that all odd harmonics vanish). In the case of 
BFKL linear approximation, we have $xG_{ \textrm{DP}  }(x,q_{\perp }, \Delta_\perp)
=- (q_\perp^2 -\Delta_\perp^2/4)\frac{2N_c}{\alpha_s}  \mathcal{T}(q_\perp,\Delta_\perp, Y) $ 
for the case with finite momentum transfer.

{\it Correlated Hard Diffractive Dijet Production in DIS.}
\begin{figure}[tbp]
\begin{center}
\includegraphics[width=7cm]{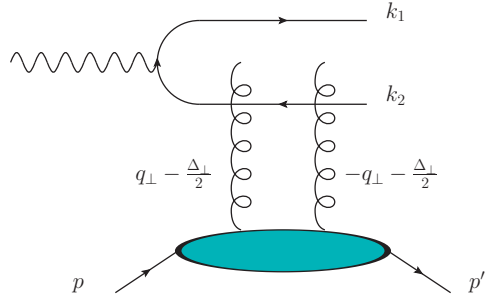}
\end{center}
\caption[*]{Diffractive dijet production in electron-ion collisions. Here we assume that the 
incoming virtual photon has only the longitudinal momentum. The signature of the 
diffractive process is the rapidity gap between the produced dijet and the target hadron which remains intact. }
\label{a}
\end{figure}
Now let us discuss diffractive dijet production in electron-ion collisions, which has 
been studied quite recently in Ref.~\cite{Altinoluk:2015dpi}, and demonstrate that 
it directly probes the dipole gluon GTMD. Diffractive events imply that a color 
neutral exchange must occur in the $t$-channel between the virtual photon and 
the target hadron over several units in rapidity. Following the same framework 
developed in Ref~\cite{Dominguez:2011wm}, by requiring that the final state 
quark-antiquark pair forms a color singlet state, we can write the cross section 
for diffractive dijet production as illustrated in Fig.~\ref{a} as follows
\begin{eqnarray}
\frac{d\sigma ^{\gamma_{T}^{\ast }A\rightarrow q\bar{q}X}}{dy_1d^2k_{1\perp}dy_2d^2k_{2\perp}}
&=&2N_{c}\alpha _{em}e_{q}^{2}\delta(x_{\gamma^*}-1) z(1-z) [z^2+(1-z)^2] \int d^2 q_\perp d^2 q_\perp^\prime  \mathcal{F}_x(q_\perp, \Delta_\perp)\mathcal{F}_x(q_\perp^\prime , \Delta_\perp) \notag \\
&&\times \left[  \frac{P_\perp}{P_\perp^2 +\epsilon_f^2 } - \frac{P_\perp-q_\perp }{(P_\perp-q_\perp)^2 +\epsilon_f^2} \right] \cdot  \left[  \frac{P_\perp}{P_\perp^2 +\epsilon_f^2 }  - \frac{P_\perp-q_\perp^\prime}{(P_\perp-q_\perp^\prime)^2 +\epsilon_f^2 } \right]   \ ,\label{dct} 
\end{eqnarray}
for the transversely polarized photon. A similar cross section formula  
can be written for the longitudinally polarized photon. In Eq.~(\ref{dct}), $y_{1,2}$ and $k_{1,2\perp}$ are rapidities and
transverse momenta of the final state quark and antiquark jets, respectively, 
defined in the center of mass frame of the incoming photon and nucleon.  $\vec{P}_\perp \equiv \frac{1}{2} (\vec{k}_{2\perp}-\vec{k}_{1\perp})$  
represents the typical dijet transverse momentum and $\Delta_\perp$ 
is the nucleon recoiled momentum which equals to $-(k_{1\perp}+k_{2\perp})$ at leading order.  
We are interested in the back-to-back kinematic region for the two final state jets where 
$|P_\perp| \gg |\Delta_\perp|$. 
Suppose  $\epsilon_f^2\equiv z(1-z)Q^2$ is not too large. Then we expect 
that the above $q_\perp$ integrals are dominated by the region 
$q_\perp \sim P_\perp$ and the cross sections are roughly 
proportional to $\mathcal{F}_x^2(P_\perp, \Delta_\perp)$ for 
back-to-back dijet configurations. Thus, the diffractive
dijet production will be sensitive to the correlation between
$P_\perp$ and $\Delta_\perp$ as mentioned in Ref.~\cite{Altinoluk:2015dpi}.

With the detector capability at the future EIC~\cite{Accardi:2012qut}, we will be able to identify both
$\vec{P}_\perp$ and $\vec{\Delta}_\perp$ and measure the angular correlation
between them. In particular, the elliptic angular correlation  
$\langle \cos 2 \left(\phi_{P_\perp} -\phi_{\Delta_\perp}\right)\rangle$
can be observed in this process. 
By plugging in the $\cos2\phi$ asymmetry in the gluon Wigner distributions
from the numerical studies of Balitsky-Kovchegov evolution with impact parameter 
dependence~\cite{GolecBiernat:2003ym,Berger:2011ew}, we find that it will lead to 
a few percent $\langle \cos 2 \left(\phi_{P_\perp} -\phi_{\Delta_\perp}\right)\rangle$ 
asymmetries in the typical EIC kinematics. More sophistic calculations 
shall follow to generalize the saturation models~\cite{Kowalski:2003hm, Iancu:2003ge, Watt:2007nr, Rezaeian:2012ji} 
to incorporate this particular angular correlation feature. We leave that for a future
study. Comparing the theoretical computations with the future experimental
data will  provide us much more insights on the experimental 
signature of small-$x$ dynamics. 

{\it Summary and Discussions.} To conclude, let us make some further but brief 
comments on the consequence of this work, while we will leave the detailed 
discussion for a future publication.
\begin{itemize}
\item In principle, as far as the gluon Wigner distribution is concerned, there 
should be correlation between the two vectors $\vec{q}_\perp$ and 
$\vec{b}_\perp$, which can be shown in theoretical studies for the 
dipole scattering amplitude in the small-$x$ region. In order to 
demonstrate this non-trivial correlation, we parametrize the
above Wigner distribution as
\begin{eqnarray}
xW_g^T(x,\vec{q}_\perp;\vec{b}_\perp)=x{\cal W}_g^T(x,|\vec{q}_\perp|,|\vec{b}_\perp|)+2\cos(2\phi)
x{\cal W}_g^\epsilon(x,|\vec{q}_\perp|,|\vec{b}_\perp|) +\cdots \ ,
\end{eqnarray}
where $\phi$ is the azimuthal angle between $\vec{q}_\perp$ and $\vec{b}_\perp$. The first term above
represents the azimuthal symmetric distribution, whereas the rest of other terms stand for the azimuthal
asymmetric distribution. For example, due to the $\cos(2\phi)$ nature of the second term, we call it the Elliptic Gluon
Wigner Distribution, and in short, elliptic gluon distribution. This is quite similar to the elliptic
flow phenomena observed in heavy ion collisions.

\item Let us further comment on the WW gluon distribution case. Following the same 
technique used above for the dipole gluon Wigner distribution, we generalize the WW 
gluon distribution at small-$x$ as follows
\begin{equation}
xG_{ \textrm{WW}  }(x,q_{\perp }, \Delta_\perp)=2\int \frac{\text{d}\xi^{-}\text{d}%
^{2}\xi _{\perp }e^{-iq_{\perp }\cdot \xi _{\perp }-ixP^{+}\xi ^{-}}}{\left(
2\pi \right) ^{3}P^{+}}\left\langle  P+\frac{ \Delta_\perp}{2}\left|\text{Tr}\left[ F\left(\frac{\xi}{2}\right)
\mathcal{U}^{\left[ +\right] \dagger }F\left( -\frac{\xi}{2}\right) \mathcal{U}^{\left[ +%
\right] }\right] \right| P-\frac{ \Delta_\perp}{2}\right\rangle , \label{gtd2}
\end{equation}
which allows us to find
\begin{eqnarray}
xG_{\textrm{WW}}(x,q_\perp, \Delta_\perp)&=&\frac{2N_c}{\alpha_S}\int\frac{d^2R_\perp}{(2\pi)^2}\frac{d^2R_\perp^\prime}{(2\pi)^2}\;e^{iq_\perp\cdot(R_\perp-R_\perp^\prime)+i \frac{\Delta_\perp}{2}\cdot(R_\perp+R_\perp^\prime)}\notag \\
  &&\times \frac{1}{N_c}\left\langle\text{Tr}\left[i\partial_iU(R_\perp)\right]U^\dagger(R_\perp^\prime)\left[i\partial_iU(R_\perp^\prime)\right]U^\dagger(R_\perp)\right\rangle_{x}.
\end{eqnarray}
Due to the known connection between the WW gluon distribution and color quadrupoles at small-$x$~\cite{Dominguez:2011wm}, 
it is expected that one needs to generate a color quadrupole at the amplitude level in order to probe the WW Wigner distribution. 
This requires two incoming photons at once which produce four-jet diffractive events in the final states. It seems to be very 
challenging to measure this type of events at EIC. Nevertheless, it is more probable to perform such measurement in 
ultra-peripheral diffractive $AA$ collisions at the LHC where photons are much more abundant in the wavefunction of colliding nuclei. 

\item
It is also interesting to note that one can generalize the above derivation to obtain the linearly polarized 
part~\cite{Mulders:2000sh, Boer:2010zf, Qiu:2011ai, Metz:2011wb, Dominguez:2011br, Sun:2011iw, 
Boer:2011kf, Dumitru:2015gaa} of the WW and dipole gluon Wigner distribution when the indices of 
derivatives are off-diagonal, instead of diagonal as in Eqs.~(1,2). The cross sections for dijet and 
four-jet productions depend on both the unpolarized and linearly polarized gluon distributions, 
which are related in the small-$x$ formalism~\cite{Metz:2011wb, Dominguez:2011br}.

In addition, when integrating over $q_\perp$ in Eqs.~(1,2) with off-diagonal indices, the gluon Wigner
distributions will reduce to the so-called helicity flip gluon GPDs (also called gluon transveristy), which 
have been extensively discussed in the collinear GPD framework~\cite{Hoodbhoy:1998vm,Belitsky:2000jk,Diehl:2001pm,Belitsky:2001ns}. 
The nontrivial correlations between $q_\perp$ and $\Delta_\perp$ play important roles 
in the integral to obtain the helicity flip gluon GPDs.
\end{itemize}

The parton Wigner distributions, which contain the most complete information, are the cornerstones of 
all parton distributions. We demonstrate that gluon Wigner distributions are closely related to the 
impact parameter dependent dipole and quadrupole scattering amplitudes, and point out that they 
can be measured in diffractive type events at EIC and the LHC. 

{\it Acknowledgments:}
This material is based upon work supported by the U.S. Department of Energy,
Office of Science, Office of Nuclear Physics, under contract number
DE-AC02-05CH11231, and by the U.S. National
Science Foundation under Grant No. PHY-0855561 and PHY-1417326. B.X. acknowledges interesting discussion with A. Mueller, and wishes to thank Dr. X.N. Wang and the nuclear theory group at LBNL for hospitality and support during his visit when this work is finalized.

\end{document}